\documentclass[showpacs,amsmath,amssymb,floatfix,
superscriptaddress,notitlepage,iopart]{revtex4-1}

\usepackage{color}
\usepackage{bbm}
\usepackage{graphicx}
\usepackage{dcolumn}
\usepackage[utf8]{inputenc}
\usepackage{graphicx}
\usepackage{epstopdf}
\usepackage{amsthm}
\usepackage{amsmath}
\usepackage{empheq}
\usepackage{bbm}
\usepackage{braket}
\usepackage{amssymb}
\usepackage[usenames,dvipsnames]{xcolor}
\usepackage{cases}
\usepackage{latexsym}
\usepackage[colorlinks=true,citecolor=Cerulean,linkcolor=RubineRed,urlcolor=Cerulean]{hyperref}

\graphicspath{ {images/} }

\newcommand{\id}{\mathbbm{1}} 
\newcommand{\tr}[1]{\operatorname{\textnormal{Tr}}\left( {#1} \right)}  
\newcommand{\vel}{\operatorname{\mathcal{V}}} 
\newcommand{\rhog}[1]{ {\rho_{#1}^\mathcal{C}}} 
\newcommand{\xg}[1]{ {x_{#1}^\mathcal{C}}} 
\newcommand{\yg}[1]{ {y_{#1}^\mathcal{C}}} 
\newcommand{\zg}[1]{ {z_{#1}^\mathcal{C}}} 
\newcommand{\velg}{\operatorname{\mathcal{V}_\mathcal{C}}} 

\usepackage{graphicx}

\begin{document}

\title{Quantum Speed Limits under Continuous Quantum Measurements}
\author{Luis Pedro Garc\'ia-Pintos}
\affiliation{Department of Physics, University of Massachusetts, Boston, MA 02125, USA}
\author{Adolfo del Campo}
\affiliation{Department of Physics, University of Massachusetts, Boston, MA 02125, USA}
\affiliation{Theory Division, Los Alamos National Laboratory, MS-B213, Los Alamos, NM 87545, USA}
\date{\today}

\begin{abstract}
The pace of evolution of physical systems is fundamentally constrained by quantum speed limits (QSL), which have found broad applications in quantum science and technology.
We consider the speed of evolution for quantum systems undergoing stochastic dynamics due to continuous measurements. It is shown that that there are trajectories for which  standard QSL are violated, and we provide estimates for the range of velocities in an ensemble of realizations of continuous measurement records. We determine the dispersion of the speed of evolution and characterize the full statistics of single trajectories. By characterizing the dispersion of the Bures angle, we further show that continuous quantum measurements induce Brownian dynamics in Hilbert space.
\end{abstract}

\maketitle

\section{Introduction}

By definition, a physical process  concerns the evolution of a physical system of interest. The quest for elucidating the limits of the underlying dynamics has thus proved useful across a wide variety of fields beyond the realm of nonequilibrium physics. 
Progress to this end has been guided by the use of time-energy uncertainty relations both in the classical and quantum domain. In the later case, it was soon understood that fundamental limits to quantum dynamics could be sharpened by analyzing the notion of passage time, this is, the time required for a quantum state to evolve into an orthogonal state \cite{booktimevol1,booktimevol2}.
Building on the seminal work by Mandelstam and Tamm \cite{mandelstamtamm1945}, Quantum Speed Limits (QSL) have been formulated to provide a lower bound on the passage time, or more generally, on the minimum time required for a state to evolve into a distinguishable state under a given dynamics.
It is by now understood that QSL impose constraints on parameter estimation in quantum metrology \cite{demkowicz2012NatComm,BeauPRL2017} and on quantum control protocols \cite{GiovanettiPRL2009,AdCPRL2012,An2016,DS17,DeffnerPRL2017,Funo17}. In addition, they can be used to ascertain the ultimate computational power of physical systems~\cite{margoluslevitin1998,LloydNature2000,LloydPRL2002,MacconePRA2003} and  the performance of thermodynamic devices such as quantum engines and batteries \cite{delcampo14,Campaioli17}. 
More recently, QSL have been 
connected to quantum coherence resource theory~\cite{MarvianPRA2016,PlenioRevModPhys2017}. Further, it has been  
established that the existence of speed limits is not restricted to the quantum realm, and that similar bounds to the speed of evolution can be found for classical processes as well~\cite{Shanahan18,Okuyama18}.

QSL have been generalized to systems embedded in an environment, when the dynamics is open \cite{DavidovichPRL2013,delCampoPRL2013,DeffnerLutzPRL2013,Marvian15,Chenu17,Beau17decay}. Yet, an important class of open quantum dynamics results 
from the monitoring  of a system by an observer. The  dynamics is then described by continuous quantum measurements \cite{MenskyPRD1979, DiosiPhysLettA1986,
CavesPRD1986,
DiosiPhysLettA1988,
BelavkingJPA1991,
MenskyPhysLettA1994,Wiseman1996,
KorotkovPRB2001} and cannot be accommodated by the aforementioned results  \cite{DavidovichPRL2013,delCampoPRL2013,DeffnerLutzPRL2013,Marvian15,Chenu17,Beau17decay} as the equation of motion becomes nonlinear on the quantum state. Continuous quantum measurements  can be understood as the result of a  sequence of infinitesimally weak measurements and  are  natural in physical settings in which the system-measurement device coupling is weak~\cite{MurchNature2013,SiddiqiNature2014,HuardPRX2016}. 
As such, and in parallel to QSL, they have given rise to advances in parameter estimation~\cite{VerstraetePRA2001,SmithPRL2006,ChasePRA2009,
TsangPRA2009,WheatleyPRL2010,RalphPRA2011,WangPRL2015,
MolmerPRA2016,CortezPRA2017,AlbarelliNJP2017}, quantum control~\cite{DohertyPRA1999,AhnPRA2002,AhnPRA2003,JacobsPRL2007,CombesPRA2010,RisteNat2013,HacohenPRL2018}, and foundations of physics \cite{ScienceDevoret2013,PalaciosNatPhys2010,MurchNature2013,SiddiqiPRL2014}. 
The ability to experimentally measure and manipulate individual quantum trajectories in this scenario motivates our study.
For non-Hermitian Hamiltonians, a related quantum jump approach has been used to show that spatial correlations can
propagate faster than the so-called Lieb-Robinson bounds~\cite{AshidaUeda18},
 while the speed of evolution in the presence of gain and loss, that is deterministic but non-linear in the quantum state, has been considered in \cite{BrodyGraefe12}

In this work, we explore the fundamental limits to the pace of evolution of a physical system under continuous quantum measurements.  We show that the speed of evolution of a continuously-monitored physical system is not subjected to known QSL.  While the later remain applicable  to the ensemble dynamics, they are violated  in individual realizations of a continuous measurement experiment.  As a result, in a single trajectory a quantum system can evolve from an initial to a distinguishable state at velocities exceeding established bounds. The distribution of velocities in an ensemble of trajectories can be highly asymmetric, with a residual fraction of trajectories at the expected QSL, which  forces a reconsideration of the precise physical meaning and implications of QSL.

\section{Methods}

{\it QSL under continuous measurements.---}
In order to study the limits on the speed of evolution of a physical system, we focus on the 
Uhlmann-Josza
fidelity between its initial state $\rho_0$ and the state $\rho_t$ at a time $t$,
given by $F(\rho_0,\rho_t)~=~[\tr{ \sqrt{ \sqrt{\rho_0} \rho_t \sqrt{\rho_0} } }]^2$ ~\cite{Uhlmann1993,jozsa1994fidelity}. While the fidelity is not a metric, it provides a notion of proximity between the  two states ~\cite{Booknielsenchuang}. 
We shall be interested in the dynamics of a quantum system initialized in a pure state. In this case the  fidelity reads
\begin{align}
    F(t) = \tr{\rho_0 \rho_t}.
\end{align}
We note that other information-theoretic quantifiers of the distinguishability of the states $\rho_0$ and $\rho_t$ can be considered \cite{Pires16}.
When the dynamics is open, the time evolution of  the quantum state $\rho_t$ can often be described by a master equation of the form \cite{Rivas12}
\begin{align}
\label{eq:mastereq}
    \frac{d \rho_t}{dt} &=  - i [H,\rho_t] + \mathcal{D} [\rho_t] \equiv L [\rho_t], 
\end{align}
where  $H$ denotes the system Hamiltonian (with units such that $\hbar = 1$), the dissipator $\mathcal{D} [\rho_t] $ is responsible for the nonunitary character of the evolution, and $L$ is generally referred to as the Liouvillian.

The change in the fidelity from an initial value $F(0) = 1$ to some final value $F(\tau)$ after a time $\tau$ is related to the speed of evolution via
\begin{align}
\label{eq:changeF}
    \Delta F = \int_0^\tau \dot{F}(t) dt  &= \int_0^\tau \tr{\rho_0 L[\rho_t]}dt = - \tau \vel, 
\end{align}
where the time-averaged velocity is defined as 
\begin{equation}
\label{eq:velocity}
\vel = -\frac{1}{\tau} \int_0^\tau \tr{\rho_0 L[\rho_t]} dt \equiv -\overline{\tr{\rho_0 L[\rho_t]}} ,
\end{equation} 
and $\overline{ g(t) } \equiv \int_0^\tau g(t) dt/\tau$ denotes time average over the window $[0,\tau]$. Note that, since $\Delta F \le 0$, the velocity is positive.
The time necessary for a change of $\Delta F$ in the fidelity is then  $\tau =-\Delta F/ \vel$.
Rewriting the  fidelity  as $F = \text{cos}^2(\mathcal{L})$ in terms of   the Bures angle $\mathcal{L}$ \cite{Uhlmann92}, which does define a metric in Hilbert space, the time necessary to sweep 
$\mathcal{L}$ is related to the speed of evolution by
\begin{align}
\label{eq:timefromvelocityBures}
    \tau &= \frac{ \text{sin}^2\mathcal{L}(\tau) }{ \mathcal{V} }.
\end{align}
QSL result from finding upper bounds to the speed of evolution $ \mathcal{V}$, i.e., lower bounds to $\tau$.

For the open dynamics described by the master equation (\ref{eq:mastereq}), the average velocity $\vel$ can be upper bounded in terms of various norms of $ L[\rho_t]$, directly leading to a lower bound on the time $\tau$ necessary for the system to reach a target state ~\cite{DavidovichPRL2013,delCampoPRL2013,DeffnerLutzPRL2013}.
The tightest QSL for open system dynamics comes from bounding the velocity by the operator norm of the Liouvillian, in which case one finds that $\vel \le \vel_{\rm QSL}$, with
\begin{align}
\label{eq:velQSL}
\vel_{\rm QSL} = \frac{1}{\tau} \int_0^\tau \| L(\rho_t) \|,
\end{align}
where $\| \cdot \|$ denotes the operator norm, i.e. the largest singular value~\cite{DeffnerLutzPRL2013}.

Of particular relevance to our following analysis is Markovian dynamics, where the standard Lindblad form of $L(\rho_t)$ simplifies to
\begin{align}
\label{Ldepahsing}
L(\rho_t)=- i [H,\rho_t] -\sum_j \kappa_j [A_j,[A_j,\rho_t]] 
\end{align}
when the Lindblad operators $A_j$ are Hermitian; $\kappa_j$ being the corresponding coefficients.  

Markovian dynamics of the form~\eqref{eq:mastereq} and~\eqref{Ldepahsing} can arise from very different physical processes~\cite{Wiseman2001unravellings}, including 
the  interaction of the system with a memoryless environment \cite{Rivas12} or stochastic Hamiltonian fluctuations \cite{Chenu17}.
However, identical dynamics arise from the monitoring of the system by a continuous quantum measurement in a situation in which one has no access to the measurement outcomes~\cite{JacobsIntro2006}. 

We shall focus on  the continuous measurement of an observable $A$ performed on a quantum system (generalization to multiple observables is straightforward~\citep{JacobsIntro2006}). 
An observer conditions the state of the system to the measurement outcome $r$ obtained at a time $t$, during an interval $dt$. 
The conditioned state $\rhog{t}$, i.e., the state that the observer assigns to the system, changes according to
\begin{align}
\label{eq:changerho}
     d \rhog{t} &= L\left[\rhog{t}\right] dt + I \left[\rhog{t}\right] dW_t ,
\end{align}
when expressed in It\^o form \cite{Bookjacobs2014}. Here,  $L(\rhog{t} )$ takes the form of Eq. (\ref{Ldepahsing}) for the case of a single Lindblad operator
\begin{align}
\label{eq:Lindblad}
L(\rhog{t} ) &=-i \left[H,\rhog{t} \right] - \kappa \left( A^\dag A \rhog{t} + \rhog{t} A^\dag A - 2 A \rhog{t} A^\dag  \right).
\end{align}
The additional term 
\begin{align}
\label{eq:innovation}
I \left[\rhog{t}\right] &= \sqrt{2\kappa} \Big( A\rhog{t} + \rhog{t} A^\dag - \tr{\left( A + A^\dag \right) \rhog{t}} \rhog{t} \Big),  
\end{align}
is nonlinear in $\rhog{t}$ and accounts for the change of the quantum state due to the acquisition of information  during the measurement process. It is referred to as the innovation term, and is multiplied by $dW_t$, a zero-mean real Gaussian random variable with variance $dt$~\cite{JacobsIntro2006,Bookjacobs2014}. 

 The innovation term plays the role of formally conditioning to the observed outcome in a measurement. The mechanism is the generalization of the standard procedure that projects to the observed outcome after a strong projective measurement. 
For instance, following the postulates of quantum mechanics, when measuring an observable with a spectral decomposition $O = \sum_j \lambda_j \ket{\lambda_j} \bra{\lambda_j}$ and observing outcome $\lambda_j$, the state of the system is given by $\ket{\lambda_j}$. The second term in equation~\eqref{eq:changerho} generalizes this to the case of a weak measurement occurring at time $t$~\cite{JacobsIntro2006}.

Note that without access to the measurement output,  the quantum state that   best describes the system involves an   average over all the different possible outcomes. In such a case, the state is given by the ensemble average $\rho_t \equiv \langle \rhog{t} \rangle$ and evolves according to 
\begin{align}
\label{eq:ensdynamics}
     d \rho_t = \langle d \rhog{t} \rangle &= \left\langle L\left[\rhog{t}\right] \right\rangle dt = L\left[\rho_t\right] dt  ,
\end{align}
where $\langle \cdot \rangle$ denotes an average over the unknown possible results of the measurement, and we have used $\langle dW_t \rangle = 0$ and that the Liouvillian is linear in the state.
The ensemble-averaged state then evolves according to
the quantum master equation~\eqref{eq:mastereq}.  

\section{Results}

For an observer with no access to the measurement outcomes, the evolution is described by an open quantum dynamics, resulting from the coupling of the system to an inaccessible environment.  In this scenario, the observer would expect the standard QSL to hold, with the time for a quantum state to sweep a given Bures angle $\mathcal{L}(\tau)$  being given by Eqs.~\eqref{eq:velocity} and~\eqref{eq:timefromvelocityBures}.
Said differently, the description by the observer is consistent with the standard QSL as $\vel \le \vel_{\rm QSL}$.

However, we will show that the more complete description of the system provided by the state $\rhog{t}$ implies otherwise. Indeed, given that $\rhog{t}$ evolves in a stochastic manner, both the effective velocity and the Bures angle $\mathcal{L}_\mathcal{C}(\tau)$ covered become stochastic variables, for a fixed duration $\tau$ of the experiment. 
A similar derivation as for Eq.~\eqref{eq:changeF}, combined with~\eqref{eq:changerho}, gives
\begin{align}
\label{eq:conditionedtime}
    \tau &= \frac{\text{sin}^2(\mathcal{L}_\mathcal{C}(\tau)) }{ \velg },
\end{align}
where the \emph{conditioned velocity} is 
\begin{align}
\label{eq:conditionedvelocity}
    \velg &= -\overline{\tr{\rho_0 L[\rhog{t}]}} - \frac{1}{\tau} \int_{0}^{\tau} \tr{\rho_0 I[\rhog{t}]} dW_t .
\end{align}
This velocity involves a term similar to that appearing in Eq.~ \eqref{eq:timefromvelocityBures}
and a stochastic integral that depends on the information acquisition term $I$~\cite{Bookjacobs2014}. The latter modifies the rate of change for the more complete description of the system given by $\rhog{t}$, when the  measurement outcomes are known. 


Given that ensemble  and time averages commute, and that the ensemble-averaged evolution is linear in the state, 
Eq.~\eqref{eq:conditionedvelocity} immediately leads to
\begin{equation}
\langle \velg \rangle = \vel.
\end{equation}
That is, the ensemble-averaged velocity coincides with the one obtained from the ensemble-averaged evolution, which means that, on average, the passage time of the actual (conditioned) evolution of the system is the same as for the ensemble-averaged state. 

However, individual realizations can travel at a different rate. 
In order to see this, we use Eq.~\eqref{eq:conditionedvelocity} to expand the second moment of the conditioned velocity, i.e., 
\begin{align}
\left\langle \velg^2 \right\rangle & = \left\langle \overline{  \tr{\rho_0 L[\rhog{t}]}  }^2 \right\rangle  \\
    & \!\!\!+ \frac{2}{\tau} \left\langle \overline{  \tr{\rho_0 L[\rhog{t_1}]}  }  \int_0^\tau \text{Tr} \left( \rho_0 I [\rhog{t_2}]   \right) dW_{t_2}     \right\rangle    \nonumber \\
    &\!\!\!+ \frac{1}{\tau^2}   \left\langle \int_0^\tau\!\!\! \int_0^\tau\!\!\text{Tr} \left( \rho_0 I [\rhog{t_1}]   \right)    \text{Tr} \left( \rho_0 I[\rhog{t_2}]   \right) dW_{t_1} dW_{t_2}  \right\rangle . \nonumber
\end{align}
Note that the second term is non-zero since $\rhog{t}$ can depend on the value of the noise term at previous times.
On the other hand, in the It\^o representation any white noise term is uncorrelated from the past and present. 
The third term can thus be simplified using the fact that 
the noise is independent from the dynamics of the system at previous times~\cite{Bookwiseman2009}, and that $\langle dW_{t_1} dW_{t_2} \rangle =  \delta_{t_1,t_2} dt_1$. 
As a result, the exact expression for the variance of the conditioned velocity in terms of the innovation term reads
\begin{align}
\label{eq:varianceVelocity}
    \Delta_{\velg}^2(t) & = \left\langle \overline{  \tr{\rho_0 L[\rhog{t}]}  }^2 \right\rangle - \vel^2  \nonumber \\
        & \!\!\!+ \frac{2}{\tau} \left\langle \overline{  \tr{\rho_0 L[\rhog{t_1}]}  }  \int_0^\tau \text{Tr} \left( \rho_0 I [\rhog{t_2}]   \right) dW_{t_2}     \right\rangle    \nonumber \\
    &+\frac{1}{\tau}\left\langle \overline{  \tr{\rho_0 I[\rhog{t}]}^2  } \right\rangle.
\end{align}
Note that in the absence of conditioning to the measurement outcomes, the variance  $\Delta_{\velg}^2(t) = 0$, given that $\rhog{t} \rightarrow \rho_t$ and that the terms that depend on $I$ vanish. The velocity thus becomes deterministic in this case. 

The indeterminacy in the velocity comes solely from the possibility of giving a more complete description of the state by means of the stochastic measurement record.
Indeed, the Cauchy-Schwarz inequality implies that $\left\langle \overline{  \tr{\rho_0 L[\rhog{t}]}  }^2 \right\rangle \ge \left\langle \overline{  \tr{\rho_0 L[\rhog{t}]}  } \right\rangle^2 = \vel^2 $, which means that the velocity has a non-zero variance whenever the information acquisition term acts non-trivially.
That is, while on average the speed of evolution of the conditioned state is the same as the ensemble averaged one, some trajectories are \emph{necessarily} faster than what an observer ignorant of the outcomes would expect. 
Established QSL are therefore violated under continuous quantum measurements, as we illustrate next in the paradigmatic example of a qubit.

\emph{Continuous measurements on a qubit.---}
The experimental continuous quantum measurement of a qubit has been reported in \cite{SiddiqiNature2014,SmithPRL2006,MurchNature2013}.
Let us consider the continuous measurement of the operator $A~=~\sigma_z$ on a qubit described by the Hamiltonian
\begin{align}
H = \frac{\omega}{2}\sigma_y,
\end{align}
where $\omega$ is the driving frequency and $\{\sigma_x,\sigma_y,\sigma_z\}$ are the Pauli matrices.
The dynamics of the conditioned state $\rhog{t}$ is dictated by Eqs.~\eqref{eq:changerho}-\eqref{eq:innovation}, while that of the ensemble-averaged state $\rho_{t}$ is given by Eq.~\eqref{eq:ensdynamics}, see the Appendix for further details.

\begin{figure}[t]
\centering
 \includegraphics[width= 0.55\columnwidth]{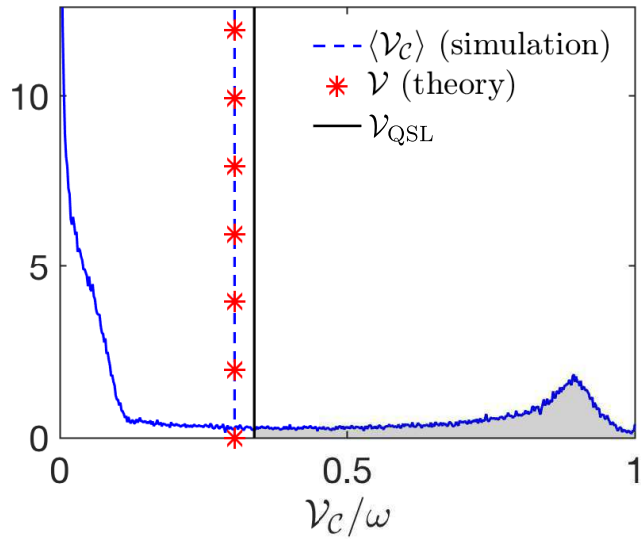}       
 \caption{\label{fig:velocityhistogram} 
 {\bf Probability distribution of the conditioned velocity $\velg$.} 
 The continuous blue line denotes the
 estimated probability distribution constructed from $10^4$ realizations of the velocity $\velg$ conditioned to a set of outcomes from the continuous measurements of $\sigma_z$ on a qubit, with a measurement constant $\kappa = \omega/4 $ and a total duration $\tau = \omega$. The black line marks the standard quantum speed limit $\vel_{\rm QSL}$, given by Eq.~\eqref{eq:velQSL}, obtained from the ensemble-averaged state that an observer with no knowledge of the measurement outcomes would assign to the system.
The shaded region illustrates the fraction of trajectories (more than $35\%$) that have a conditioned velocity which violates the ensemble-averaged quantum speed limit, $\velg > \vel_{\rm QSL}$.            }
\end{figure}

We find that the conditioned velocity of the qubit differs from $\vel$, and that,
more importantly, for some trajectories $\velg > \vel_{\rm QSL}$.
This is illustrated in Fig.~\ref{fig:velocityhistogram} by the estimated probability distribution of the conditioned velocity.
The distribution is highly asymmetric, with a large fraction of slow trajectories, and a significant fraction of the trajectories violating the standard QSL.
There exist trajectories that even approach the maximum possible velocity of $1/\tau$, that is derived from Eq.~\eqref{eq:conditionedtime} by noting that $\text{sin}^2(\mathcal{L}_\mathcal{C}(\tau)) \le 1$.   In the regime selected for Fig. \ref{fig:velocityhistogram}, $35\%$ of the trajectories exceed $\vel_{\rm QSL}$. In other regimes the violation can be even more extreme, with nearly half of the trajectories traveling faster than allowed by the ensemble evolution, as shown by further numerical simulations in the Appendix.

Remarkably, the probability distribution is not peaked at $\vel_{\rm QSL}$. Instead, only a small fraction of trajectories have velocities close to this value.
This can be understood from the behavior of the individual trajectories of the qubit under continuous monitoring of $A~=~\sigma_z$. While the effect of the unitary evolution is to rotate the qubit around the $x-z$ plane, the measurement process breaks the symmetry, favoring the poles $\pm z$ of the Bloch sphere as attractors of the evolution. 
The conditioned velocity 
inherits this behavior and takes values corresponding to the qubit having spent more time around one pole or the other, leading to a typically bimodal distribution.

\begin{figure}
 \centering
 \includegraphics[width= 0.55\columnwidth]{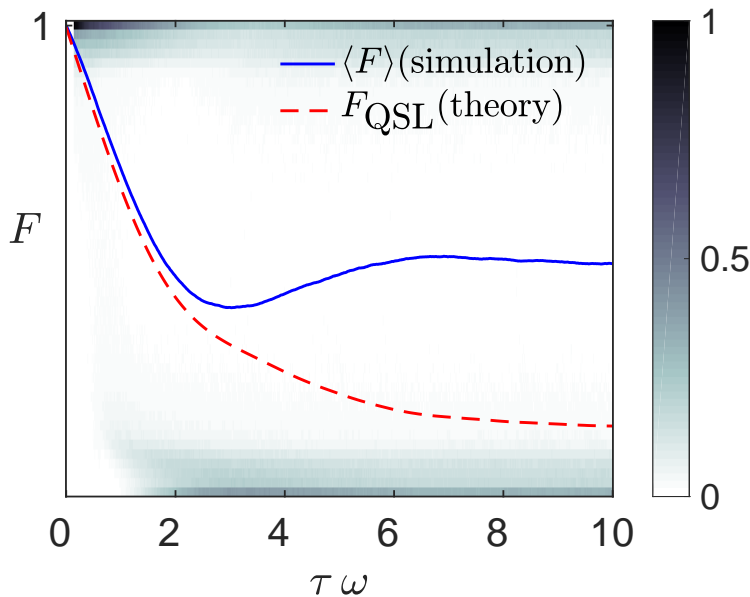} \hspace{-0.0cm} 
 \caption{\label{fig:fidelity}
{\bf Fidelity decay under continuous measurements.}     
Colormap of the fidelity as a function of time, for $\kappa = \omega/4 $ obtained from $10^4$ realizations. 
The fidelity achieved by the ensemble average (blue line) always falls below the achievable fidelity $F_{\rm QSL}$ by a system traveling at the QSL velocity (red dashed line). However, the colormap illustrates a high fraction of trajectories traversing to the forbidden region, and reaching fidelities lower than $F_{\rm QSL}$, more than $40\%$ for the final time.  }
\end{figure}

This behavior can also be seen in Fig.~\ref{fig:fidelity}, which provides a density plot comparing the fidelity decay of individual trajectories and of the ensemble-averaged dynamics as a function of time.
While the ensemble-averaged dynamics is slower than the QSL, a fraction of individual trajectories violate it at any given time. 
This leads to a large fraction of trajectories being found in a region that is forbidden to any evolution satisfying established QSL.
An additional example in which the quantum speed limit is achieved by the ensemble dynamics for a wide range of times, with $\vel$ saturating the bound $\vel_{\rm QSL}$, can be found in the Appendix.

\begin{figure}[t]
\centering
\includegraphics[width= 0.39\columnwidth]{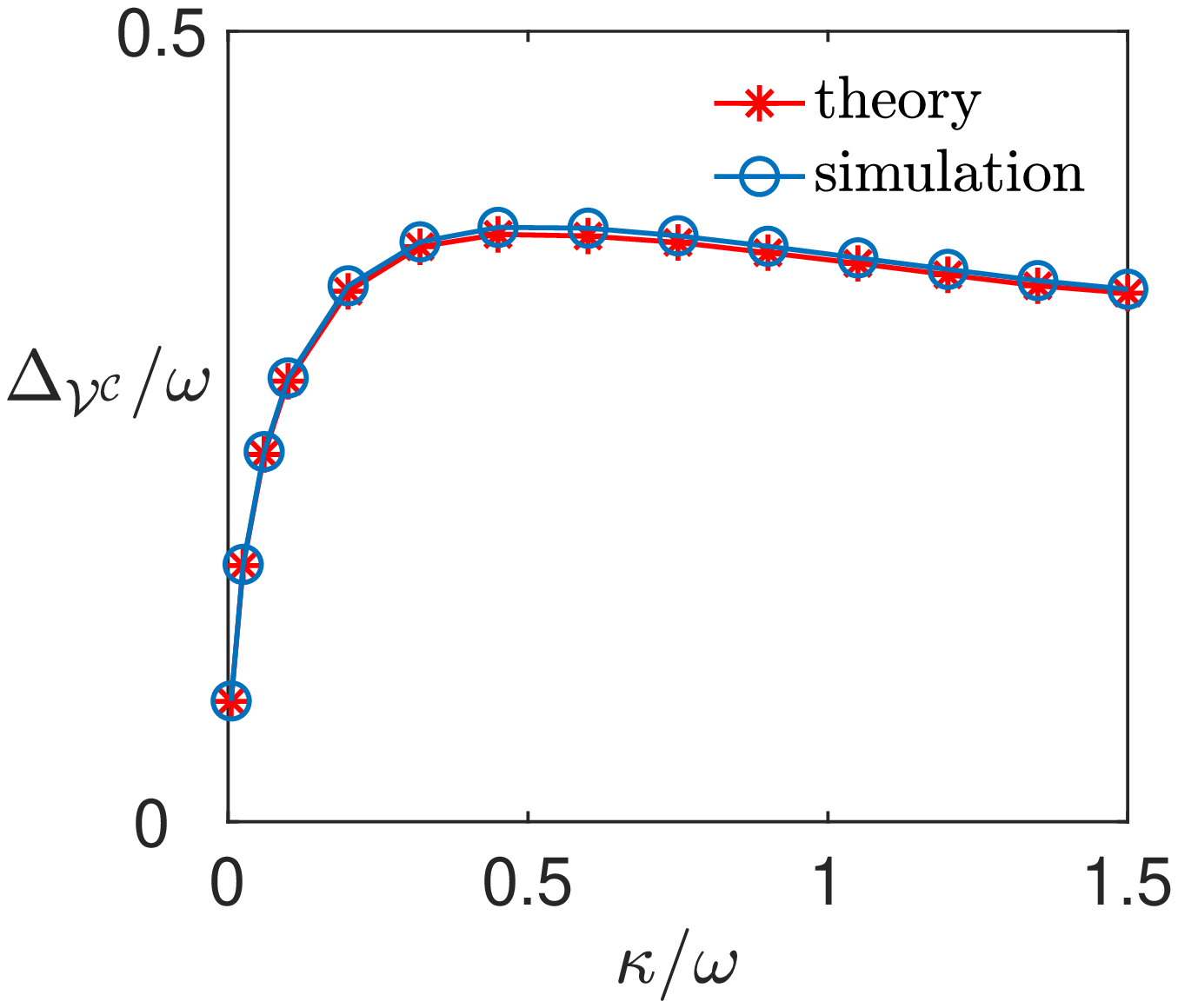} 
 \includegraphics[width= 0.45\columnwidth]{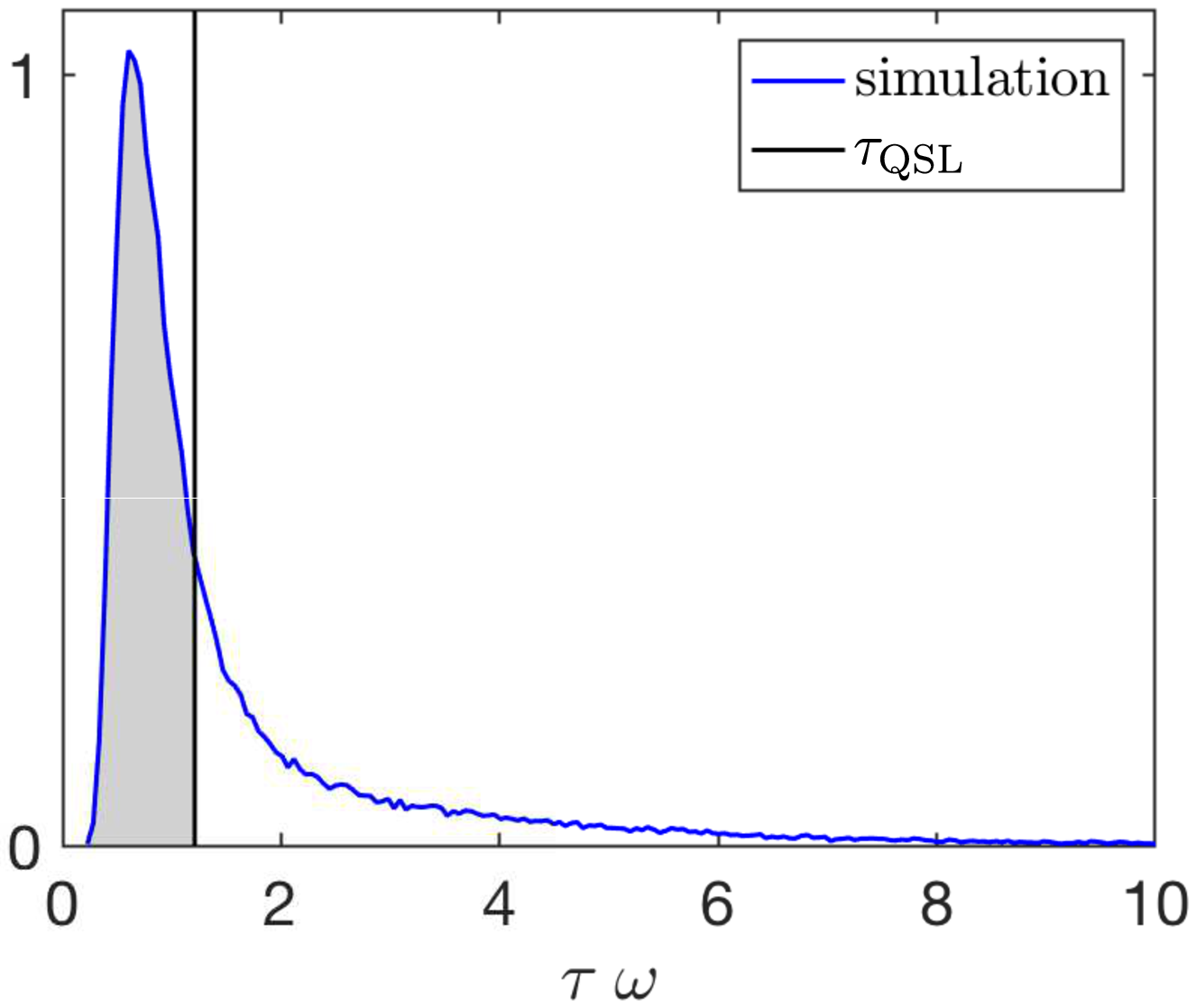} 
\caption{\label{fig:velocityfluctuationsandtimehistogram}
{\bf Velocity standard deviation vs measurement strength  and distribution of passage times.}  
Left: Standard deviation $\Delta_{\velg}$ of the conditioned velocity as a function of the measurement strength $\kappa$, for $\tau = 1/\omega$. For weak measurements ($\kappa \ll \omega$) the velocity has small uncertainty, in which case violations of the QSL become unlikely. As the measurement becomes more invasive uncertainty increases, and so do the possible violations of the QSL. 
Right: Normalized probability distribution constructed from $10^5$ realizations for the passage time $\tau$ required to reach a fixed target Bures angle $\mathcal{L} = \pi/4$, conditioned to a set of outcomes from the continuous measurements of $\sigma_z$ on a qubit with a measurement constant $\kappa = \omega/4$. 
The black line marks the time $\tau_{\rm QSL}$ necessary for the ensemble-averaged state, assigned to the system by an observer ignorant of the measurement outcomes, to reach $\mathcal{L} = \pi/4$.
The shaded region illustrates the fraction of trajectories (more than $38\%$) that have reached the target Bures angle in a time shorter than that expected by standard QSLs derived from the ensemble-averaged dynamics.
}
\end{figure}

A coarse visualization of the amount by which trajectories violate QSL is given by the standard deviation $\Delta_{\velg}$ of the conditioned velocity, which we illustrate in the left panel of Fig.~\ref{fig:velocityfluctuationsandtimehistogram} as a function of the measurement strength.
For weak measurement strength, with $\kappa / \omega \ll 1$, the standard deviation $\Delta_{\velg}$ is small. This is to be expected given that, in such a case, the conditioning to the measurement results does not considerably affect the qubit dynamics.
On the other hand, when $\kappa$ becomes comparable to the drive frequency $\omega$ the measurement induces a stronger back-action on the qubit dynamics, which now further deviates from the master equation~\eqref{eq:ensdynamics}. As a result, fluctuations in the conditioned velocity increase, with increasing violations of standard QSL. 
Naturally, such fluctuations 
directly influence the passage time of individual trajectories. This is illustrated in Fig.~\ref{fig:velocityfluctuationsandtimehistogram} (right), which depicts the distribution of times necessary for the system to travel to a fixed target Bures angle, in different realizations.
Once again, a high percentage of trajectories reach the target angle in a shorter time than $\tau_{\rm QSL}=\mathcal{L}/\mathcal{V}_{\rm QSL}$ that is expected from ensemble-averaged dynamics, thus surpassing the established QSL.

\section{Brownian dynamics in Hilbert space}

We have seen that continuous quantum measurements broaden the distribution of velocities allowing for the violation of conventional QSL at the level of single trajectories.
In this section we show that the continuous quantum measurements induce Brownian dynamics in (projective) Hilbert space.
To this end, we start form the observation that QSL can be understood as bounds on the minimum time required to sweep a given distance in Hilbert space. 
For the latter, we continue using  the Bures angle $\mathcal{L}(\rho_0,\rho_t)=\arccos\sqrt{F(\rho_0,\rho_t)}$,
defined in terms of the fidelity $F(\rho_0,\rho_t)$ between the initial and the time-evolving states, denoted by $\rho_0$ and $\rho_t$ respectively. The Bures angle between neighbouring states $\rho$ and $\rho+d\rho$ can be used to define the line element \cite{BC94,BCM96}
\begin{equation}
\label{dL2}
d\mathcal{L}^2={\rm Tr}\left[d\rho \mathcal{R}_{\rho}^{-1}(d\rho)\right]
\end{equation}
where superoperator $ \mathcal{R}_{\rho}^{-1}(d\rho)$ admits the following expansion in the eigenbasis of $\rho = \sum_j p_j \ket{j} \bra{j}$
\begin{equation}
\mathcal{R}_{\rho}^{-1}(d\rho)=\sum_{\{j,k|p_j+p_k\neq 0\}}2\frac{\langle j|d \rho|k\rangle}{p_j+p_k}|j\rangle \langle k|.
\end{equation}
Equation (\ref{dL2}) can therefore be written in the compact form
\begin{equation}
d\mathcal{L}^2=\sum_{\{j,k|p_j+p_k\neq 0\}}2\frac{|\langle j|d\rho|k\rangle|^2}{p_j+p_k}.
\end{equation}
In what follows we focus on the infinitesimal Bures angle swept by the evolution of the conditional density matrix according to the nonlinear  stochastic master equation~\eqref{eq:changerho}, associated with the continuous monitoring of a Hermitian observable satisfying $A=A^\dag$. The line element becomes
\begin{equation}
d\mathcal{L}^2=\sum_{\{j,k|p_j+p_k\neq 0\}}2\frac{|\langle j| \left(L\left[\rhog{t}\right] dt + I \left[\rhog{t}\right] dW_t\right)|k\rangle|^2}{p_j+p_k}.
\end{equation}
Using equation~\eqref{eq:innovation} and the identity $(dW_t)^2=dt$, it is found that to leading order in $dt$,
\begin{equation}
d\mathcal{L}^2=\sum_{\{j,k|p_j+p_k\neq 0\}}4\kappa dt(p_j+p_k)[A_{jk}-\tr{\rhog{t}A}\delta_{jk}]^2,
\end{equation}
where $A_{jk}=\langle j|A|k\rangle$ is the expectation value of the observable $A$ at time $t$.
The line element is therefore given by
\begin{equation}
\label{HilbertBrown}
d\mathcal{L}^2=8\kappa {\rm Var}(A) dt,
\end{equation}
where ${\rm Var} (A)={\rm Tr}[\rhog{t} \left (A-\tr{\rhog{t}A} \right)^2]$.  
Equation (\ref{HilbertBrown}) is the main equation of this section. It states that the dispersion of the Bures angle is proportional to the time of evolution under continuous quantum measurements. This is in sharp contrast with the unitary case in which the leading term is proportional to $dt^2$.
Equation (\ref{HilbertBrown}) is the analogue of the well-known second moment  of the position $x$ in Einstein's theory of Brownian motion that grows linearly in time according to ${\rm Var}[x]=2Dt$, for a particle with diffusivity $D$.
Thus, $4\kappa {\rm Var}(A)$  plays the role of the diffusivity in Hilbert space. Note however that this quantity depends on the state of the system and varies as a function of time.

\section{Discussion and conclusion}
Quantum speed limits set an upper bound to the speed of evolution of any physical process. 
However, when the system of interest is monitored via a continuous quantum measurement, its evolution becomes stochastic and the speed of evolution depends on the record of measurement outcomes.
In such a setting, established quantum speed limits need not apply and can be violated.
Specifically, an observer with no access to the measurement outcomes of a continuous quantum measurement experiment attributes an open dynamics to the system, as if it were in contact with an external environment. The dynamics accessible to the observer is then ruled by known quantum speed limits.
By contrast,  the evolution of the actual state of the system obtained by conditioning to the measurement results violates established quantum speed limits.  In the case of a monitored qubit, the  velocity distribution extracted from the statistics of individual trajectories  is not peaked at the ensemble mean value and can be highly bimodal, with a residual fraction obeying the standard quantum speed limit.
While this does not invalidate speed limits previously considered, it forces to reconsider their regime of validity and implications. 
Our results  provide a fundamental insight on the speed of quantum evolution of monitored systems and should find broad applications including quantum control and precision sensing.

\section{Acknowledgements}
It is a pleasure to thank Fernando J. G\'omez-Ruiz, Diego Tielas, and Zhenyu Xu for insightful discussions. Funding support from the John Templeton Foundation and UMass Boston (project P20150000029279) is further acknowledged.

\bibliography{referencesQSL}

\newpage 

\section*{Appendix --- continuous measurements on a qubit}
\label{app:qubitmodel}

The dynamics of the conditioned state $\rhog{t}$ is dictated by 
\begin{align}
d \rhog{t} &= L[\rhog{t}] dt+ I [\rhog{t}] dW_t,
\end{align}
with
\begin{align}
L(\rhog{t}) &=-i\frac{\omega}{2}[\sigma_y,\rhog{t} ] - \kappa [\sigma_z,[\sigma_z,\rhog{t}]] , 
\\
I (\rhog{t}) &= \sqrt{2\kappa} \left(\{ \sigma_z,\rhog{t} \} - 2\tr{\sigma_z \rhog{t}} \rhog{t} \right).  
\end{align}

\subsection{Ensemble averaged quantities}
\label{app:qubitmodel-ensemble}

Taking an average over the unknown measurement outcomes, the ensemble-averaged state $\rho_{t}$ 
evolves according to
\begin{align}
\frac{d  \rho_t  }{dt} &= - i \frac{\omega}{2}[\sigma_y,   \rho_t  ] - \kappa [\sigma_z,[\sigma_z, \rho_t ]].
\end{align}
Expressing the state in its Bloch coordinates
\begin{align}
 \rho_t  = \frac{1}{2}\left( \id +  x_t \sigma_x +  y_t \sigma_y +  z_t \sigma_z\right),
\end{align} 
with $x_t = \tr{\sigma_x \rho_t  }$, $y_t = \tr{\sigma_y  \rho_t }$, and $z_t = \tr{\sigma_z  \rho_t }$,
the evolution is governed by the following system of differential equations,
\begin{subequations}
\begin{align}
\dot x_t  &= \omega  z_t - 4 \kappa x_t,\\
\dot y_t &= - 4 \kappa y_t, \\
\dot z_t &= - \omega  x_t.
\end{align}
\end{subequations}

The $y_t$ coordinate remains decoupled from $\{x_t,z_t\}$, and decays exponentially with a decay factor $4\kappa$. 
Meanwhile, the other coordinates satisfy the equation
\begin{align}
(\dot x_t, \dot z_t) = M (x_t,z_t),
\end{align} 
with 
\begin{equation}
 M =
\begin{bmatrix}
    -4 \kappa   & \omega \\
    -\omega   & 0
    \end{bmatrix}.
\end{equation}
The eigenvalues and eigenvectors of $M$ are 
\begin{align}
\lambda_{\pm} &= -2\kappa \pm \sqrt{4\kappa^2 - \omega^2}, \\
\vec{v}_\pm &= \left( \omega, -\lambda_\mp \right).
\end{align}
In terms of them, the evolution of the Bloch coordinates  can be written as
\begin{align}
(x_t,z_t) &= a\vec{v}_+e^{\lambda_+ t} + b\vec{v}_-e^{ \lambda_- t}  \\
&= e^{  - 2 \kappa t}\Big( a\omega e^{  \sqrt{4\kappa^2 - \omega^2} t} + b\omega e^{  -\sqrt{4\kappa^2 - \omega^2} t}, - a\lambda_+ e^{  \sqrt{4\kappa^2 - \omega^2} t} - b\lambda_- e^{ - \sqrt{4\kappa^2 - \omega^2} t} \Big), \nonumber
\end{align}
where $(a,b)$ are determined by the initial conditions $(x_0,z_0)$.
Therefore, the ensemble-averaged dynamics is given by
\begin{subequations}
\begin{align}
x_t &= e^{  - 2 \kappa t} \Big( a\omega e^{  \sqrt{4\kappa^2 - \omega^2} t} + b\omega  e^{  -\sqrt{4\kappa^2 - \omega^2} t} \Big),\\
y_t &= y_0 e^{  - 4 \kappa t}, \\
z_t &=  - e^{  - 2 \kappa t} \Big(  a\lambda_+ e^{  \sqrt{4\kappa^2 - \omega^2} t} + b\lambda_- e^{ - \sqrt{4\kappa^2 - \omega^2} t} \Big).
\end{align}
\end{subequations}
From these equations,  the decay of the fidelity --and therefore the change in Bures angle-- can be calculated as a function of time.

Moreover,  the ensemble-averaged velocity is found to be given by
\begin{align}
\label{eq:appendix-qbitvelocity}
\vel &= -\left\langle \overline{ \tr{\rho_0 L[\rhog{t}]} } \right\rangle \nonumber \\
& =   i\frac{\omega}{2} \overline{ \tr{\rho_0 \left[ \sigma_y, \rho_t \right]} } + \kappa \overline{ \tr{\rho_0 \left[ \sigma_z,\left[\sigma_z,  \rho_t \right]\right]} } \nonumber \\
&= -\frac{\omega}{2}\left( x_0 \overline{z_t} - z_0 \overline{x_t}\right) + 2\kappa\left(x_0 \overline{x_t} + y_0 \overline{y_t} \right).
\end{align}

Finally, the  quantum speed limit defined via the operator norm is given by
\begin{align}
\vel_{\rm QSL} &= \frac{1}{\tau} \int_0^\tau dt\left\| L(\rho_t) \right\| \\
&=\frac{1}{\tau} \int_0^\tau dt\left\| -i\frac{\omega}{2} [\sigma_y,\rho_t]- \kappa \left[ \sigma_z[\sigma_z,\rho_t] \right] \right\| \nonumber \\
&= \frac{1}{\tau} \int_0^\tau dt\left\| \frac{\omega}{2} \left( z_t\sigma_x - x_t\sigma_z \right) - 2\kappa\left( x_t\sigma_x + y_t \sigma_y \right) \right\| \nonumber \\
&= \frac{1}{\tau} \int_0^\tau dt \sqrt{\frac{\omega^2}{4}(x_t^2+z_t^2) + 4\kappa^2(x_t^2+y_t^2) - 2 \kappa\omega x_t z_t}.    \nonumber
\end{align}

\subsection{Standard deviation of the conditioned velocity}
\label{app:qubitmodel-velocity}

Writing the conditioned state of the qubit as
\begin{align}
    \rhog{t} = \frac{1}{2} \left( \id + \xg{t}\sigma_x + \yg{t}\sigma_y + \zg{t}\sigma_z \right), 
\end{align}
we have
\begin{align}
\label{eq:appendix-auxqbitconditionedvelocity1}
\overline{ \tr{\rho_0 L[\rhog{t}]} }    = -\frac{\omega}{2}\left( x_0 \overline{\zg{t}} - z_0 \overline{\xg{t}}\right) + 2\kappa\left( x_0 \overline{\xg{t}} + y_0 \overline{\yg{t}} \right) . \nonumber
\end{align}
In addition the following identities hold
\begin{align}
    \tr{\sigma_z \rho_0 \rhog{t}} &= \frac{ \left( z_0 + \zg{t} \right) }{2} + i\frac{ \left(x_0 \yg{t} - y_0 \xg{t} \right)}{2}, \\
        \tr{\sigma_z \rhog{t} \rho_0} &= \frac{ \left( z_0 + \zg{t} \right) }{2} + i\frac{ \left( \xg{t} y_0 - \yg{t} x_0 \right)}{2},
\end{align}
and therefore
\begin{align}
    \tr{  \rho_0 \{\sigma_z, \rho_t\}} &= z_0 + z_t ,
\end{align}
which leads to 
\begin{align}
\label{eq:appendix-auxqbitconditionedvelocity2}
    \text{Tr} \left(\rho_0 I[\rhog{t}] \right) &= \sqrt{2\kappa} \left[ \tr{\rho_0 \{ \sigma_z,\rhog{t} \}} - 2\zg{t}  \tr{\rho_0 \rhog{t}} \right] \nonumber \\ 
& = \sqrt{2\kappa} \left[ \left( z_0 + \zg{t} \right) - 2\zg{t} \tr{\rho_0 \rhog{t}} \right].
\end{align}

Thereafter, the expression for the variance of the conditioned velocity reads
\begin{align}
    \Delta_{\velg}^2(t) & = \left\langle \overline{  \tr{\rho_0 L[\rhog{t}]}  }^2 \right\rangle - \vel^2  + \frac{2}{\tau} \left\langle \overline{  \tr{\rho_0 L[\rhog{t_1}]}  } \,\,\, \int_0^\tau \text{Tr} \left( \rho_0 I [\rhog{t_2}]   \right) dW_{t_2}     \right\rangle    +\frac{1}{\tau}\left\langle \overline{  \tr{\rho_0 I[\rhog{t}]}^2  } \right\rangle.
\end{align}
Using it together with Eqs~\eqref{eq:appendix-qbitvelocity},~\eqref{eq:appendix-auxqbitconditionedvelocity1}, and~\eqref{eq:appendix-auxqbitconditionedvelocity2}, the variance of the conditioned velocity can be expressed in terms of the Bloch coordinates of the conditioned state $\rhog{t}$ and of the ensemble-averaged state $\rho_t$ as
\begin{align}
    \Delta_{\velg}^2(t) &= \left\langle \overline{  \tr{\rho_0 L[\rhog{t}]}  }^2 \right\rangle - \vel^2   + \frac{2}{\tau} \left\langle \overline{  \tr{\rho_0 L[\rhog{t_1}]}  } \,\,\, \int_0^\tau \text{Tr} \left( \rho_0 I [\rhog{t_2}]   \right) dW_{t_2}     \right\rangle    +\frac{1}{\tau}\left\langle \overline{  \tr{\rho_0 I[\rhog{t}]}^2  } \right\rangle \nonumber \\
    &=\left\langle \bigg[ \frac{\omega}{2}\left( x_0 \overline{\zg{t}} - z_0 \overline{\xg{t}}\right) - 2\kappa\left( x_0 \overline{\xg{t}} + y_0 \overline{\yg{t}} \right) \bigg]^2 \right\rangle 
    -\left[ \frac{\omega}{2}\left( x_0 \overline{z_t} - z_0 \overline{x_t}\right) - 2\kappa\left(x_0 \overline{x_t} + y_0 \overline{y_t} \right) \right]^2 \nonumber \\
    &+ \frac{2}{\tau} \Bigg\langle   \left[ -\frac{\omega}{2}\left( x_0 \overline{\zg{t}} - z_0 \overline{\xg{t}}\right) + 2\kappa\left( x_0 \overline{\xg{t}} + y_0 \overline{\yg{t}} \right) \right]                        
    \int_0^\tau    \left[ \left( z_0 + \zg{t_2} \right) - 2\zg{t_2} \tr{\rho_0 \rhog{t_2}} \right] dW_{t_2}                    \Bigg\rangle 
\nonumber \\
    &+ \frac{2\kappa}{\tau} \left\langle \overline{\left[ \left( z_0 + \zg{t} \right) - 2\zg{t} \tr{\rho_0 \rhog{t}} \right]^2}\right\rangle.
    \end{align}

\begin{figure}[!h]
  \centering
    \includegraphics[width=0.35\textwidth]{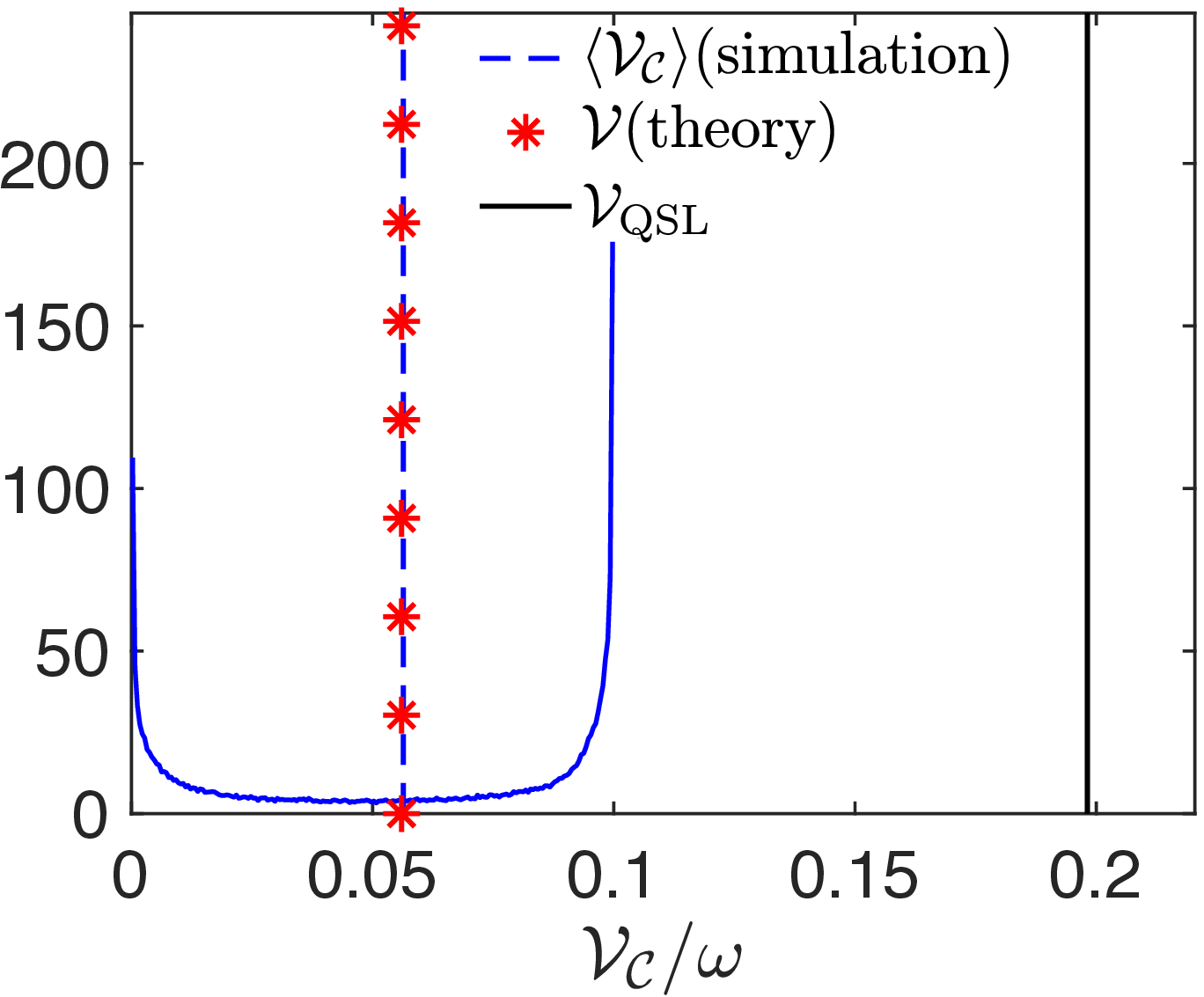}
\hspace{21pt}    
    \includegraphics[width=0.40\textwidth]{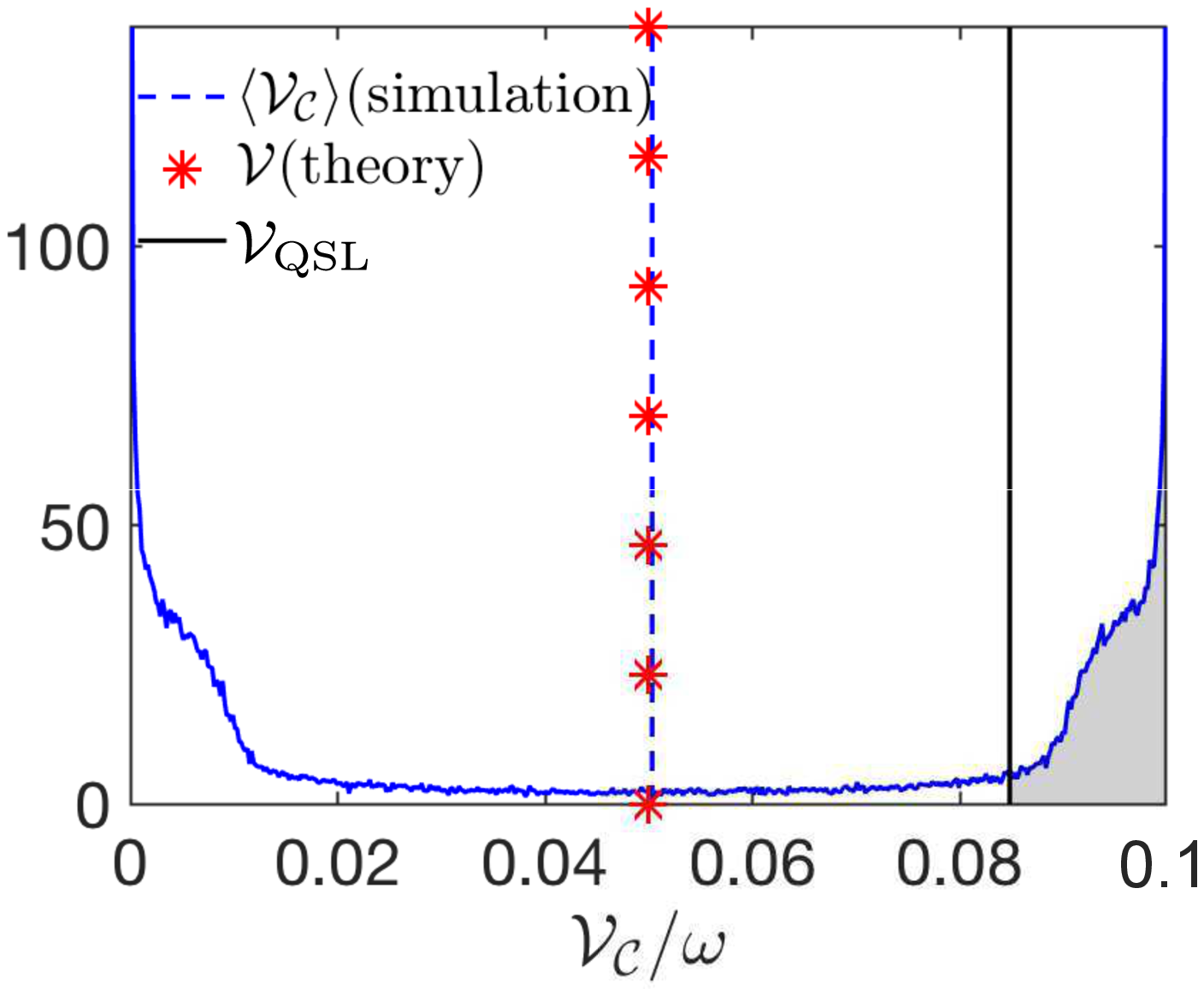} 
\\
    \includegraphics[width=0.40\textwidth]{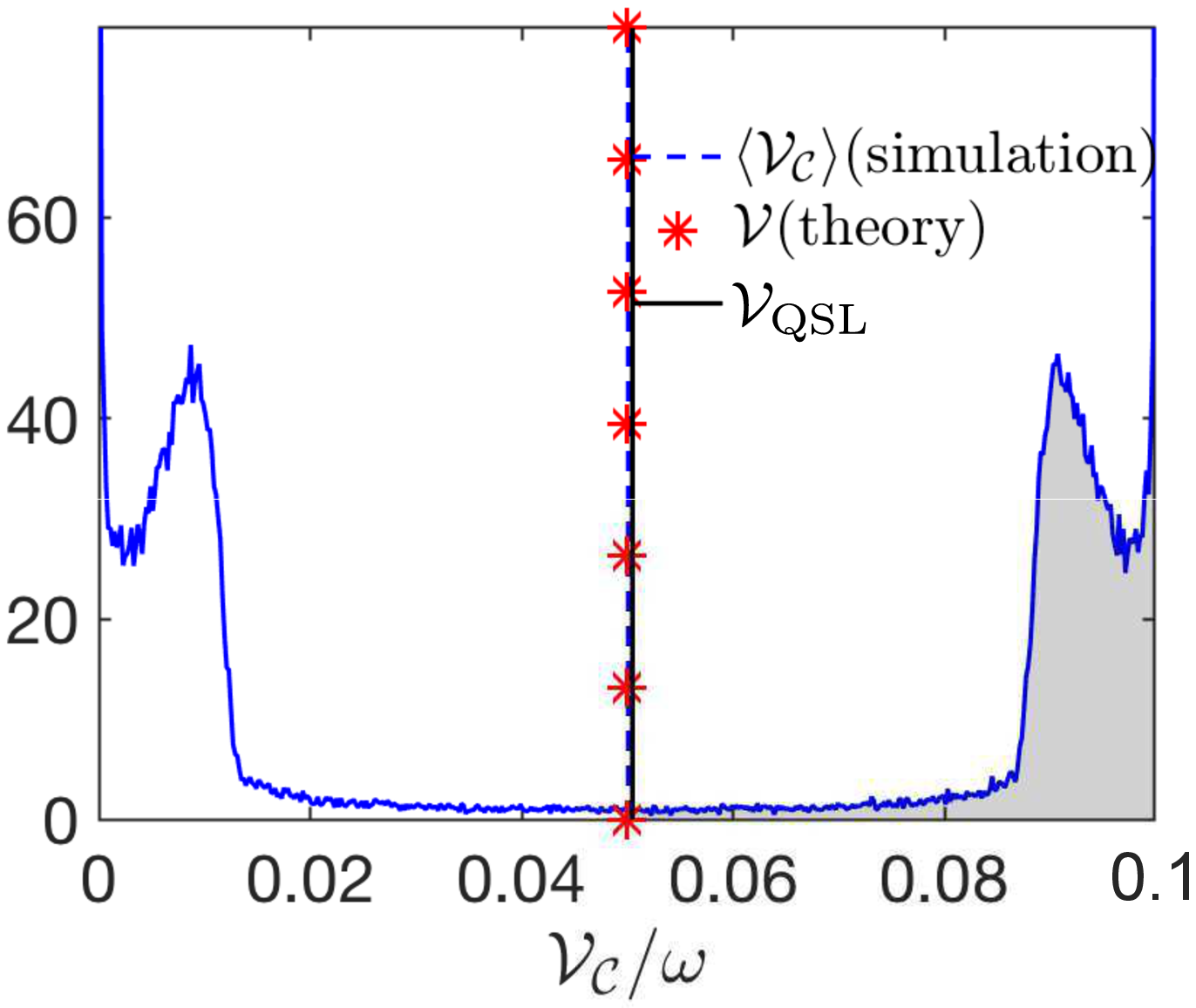}
\hspace{10pt}    
    \includegraphics[width=0.40\textwidth]{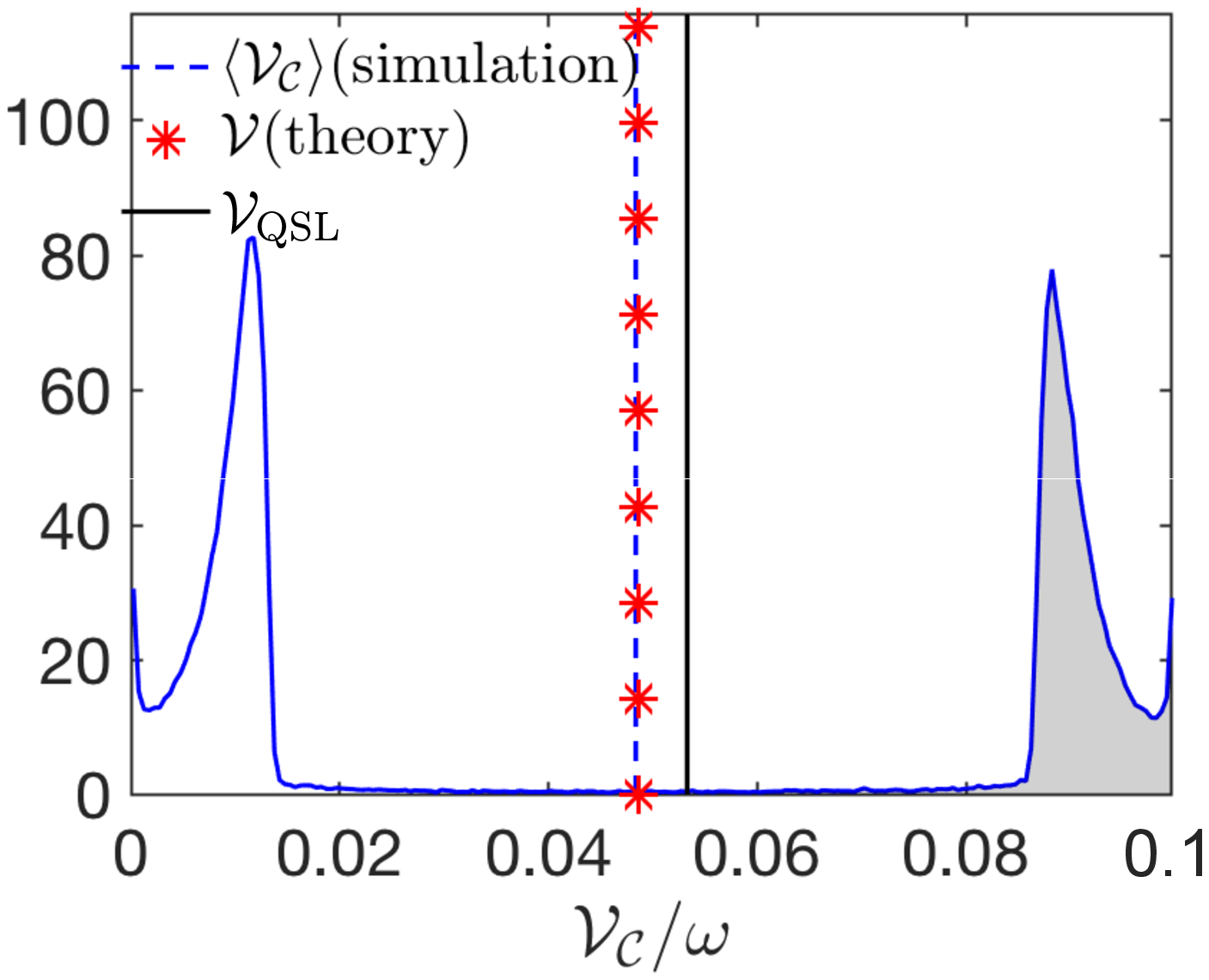}
\caption{\label{fig:app-velocityhistogram} 
{\bf Probability distribution of the conditioned velocity $\velg$.}     
The continuous blue lines denote the
approximated normalized distributions ($10^4$ realizations) of the conditioned velocity for $\tau = 10\omega$, with shaded regions illustrating the fraction of trajectories that violate the standard QSL for different values of the measurement constant $\kappa$.
While there are no trajectories with $\velg > \vel_{\rm QSL}$ for a weak measurement regime with $\kappa = 0.1 \omega$ (top left), as the measurement strength is increased trajectories begin to violate the standard bound, with
$40\%$ of trajectories violating it for $\kappa =  \omega/4$ (top right),
and about half of them violating it for $\kappa = \omega/2$ ($49\%$, bottom left),
and $\kappa = \omega$
($47\%$, bottom right).
}
\end{figure}

\label{app:qubitmodel-figures}
In this section we provide additional details on the extent to which QSL can be violated during the evolution of a qubit subject to continuous quantum measurements. To this end, we report in
Fig.~\ref{fig:app-velocityhistogram}  distributions of the conditioned velocity $\velg$ in different parameter regimes. The fraction of trajectories violating QSL are depicted in the gray shaded region. In certain regimes the distribution becomes nearly symmetric, with approximately half of the trajectories violating the QSL. The histogram is then pronouncedly peaked at slow and high velocities, with a nearly vanishing fraction of trajectories evolving at velocities near the QSL.

\begin{figure}[!h]
 \centering 
\hspace{-0.0 cm}       \includegraphics[scale=.50]{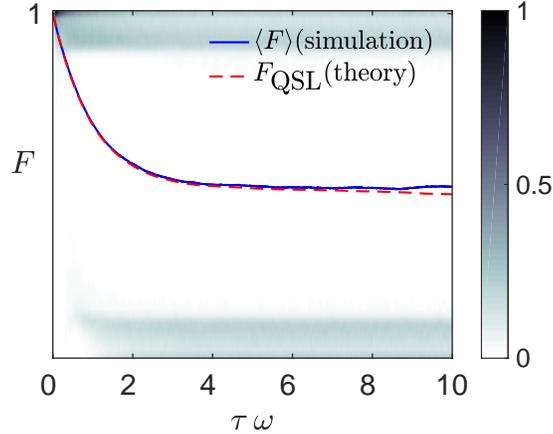} \hspace{-0.0cm} 
\caption{\label{fig:app-fidelity}
{\bf Fidelity decay vs time.}   
Colormap ($10^4$ realizations) of the Fidelity as a function of time for $\kappa = \omega/2 $. 
The Fidelity achieved by the ensemble average (blue line) approximates the achievable Fidelity $F_{\rm QSL}$ by a system traveling at the quantum speed limit velocity (red dotted line) for a wide range of times. 
However, the colormap illustrates a high fraction of particular trajectories traversing to the forbidden region, and reaching Fidelities lower than $F_{\rm QSL}$ (more than $49\%$ for the final time $\tau = 10 / \omega $).            }
\end{figure}

Meanwhile, Fig.~\ref{fig:app-fidelity} shows the fidelity decay as a function of time, comparing the ensemble-averaged dynamics with the behavior of individual trajectories.  While for short times most trajectories travel slower than the QSL, for longer times the fraction of trajectories that cross the region allowed by the QSL increases. In this example the ensemble-averaged dynamics travel at the QSL velocity, illustrating that the traditional bound $\vel_{\rm QSL}$ is tight, although only applicable for ensemble dynamics, and not for individual trajectories.

\end{document}